\documentclass[useAMS,usenatbib]{mn2e}
\usepackage{graphicx}

\usepackage{bm}
\usepackage{multirow}
\usepackage{color}
\title[Speckle Imaging with Hypertelescopes]{Speckle Imaging with Hypertelescopes}
\author[A.Surya, S.K.Saha and A.Labeyrie]{A.Surya$^{1}$\thanks{E-mail:
arun@iiap.res.in}, S.K.Saha$^{1}$, A.Labeyrie$^{2}$ \\
$^{1}$Indian Institute of Astrophysics , Bangalore\\
$^{2}$Coll\`ege de France,11, place Marcelin Berthelot,
75231 Paris Cedex 05, France}
\begin{document}

\date{Accepted . Received ; in original form }

\pagerange{\pageref{firstpage}--\pageref{lastpage}} \pubyear{2011}

\maketitle

\label{firstpage}

\begin{abstract}
Optical stellar interferometers have demonstrated milli-arcsecond resolution with few apertures spaced hundreds of meters apart. To obtain rich direct images, many  apertures will be needed, for a better sampling of the incoming wavefront. The coherent imaging thus achievable improves the sensitivity with respect to the incoherent combination of successive fringed exposures. Efficient use of highly diluted apertures for coherent imaging can be done with pupil densification, a technique also called ``hypertelescope imaging''. 
Although best done with adaptive phasing, concentrating most energy in a dominant interference  peak for a rich direct image of a complex source, such imaging is also possible with random phase errors such as caused by  turbulent ``seeing'', using methods such as speckle imaging which uses several short-exposure images to reconstruct the true image. 
We have simulated such observations using an aperture which changes through the night,  as naturally happens on Earth  with fixed grounded mirror elements, and find that reconstructed images of star clusters and extended objects are of high quality.
As part of the study we also estimated the required photon levels for achieving a good signal to noise ratio using such a technique. 
\end{abstract}


\section{INTRODUCTION}

In high-resolution optical astronomy with interferometric arrays, producing better images will require more apertures for a denser sampling of the optical wave.
As few as three apertures can suffice in principle  to reconstruct images through ``optical aperture synthesis''\citep{Baldwin}, using earth rotation or baseline changes to sample the needed
Fourier components of the object. But a better sensitivity can be reached with systems using more apertures simultaneously, even if these are smaller for a conserved collecting
area. The gain arises from the coherent combination of light vibrations achievable with many apertures working simultaneously, as opposed to the ``optical aperture synthesis'' approach
where interference fringes are recorded with fewer apertures, repeatedly with different baseline settings,  and then combined  in the computer, i.e. incoherently. N number of phased beams
combined coherently with a simple Fizeau arrangement indeed produces a highly constructive interference, in the form of a peak which is N times more intense than the average side-lobes.
Instead, successive exposures with subsets of the  sub-apertures, even if they are enlarged to conserve the photon flux and moved to improve the spatial frequency coverage, reduce the peak intensity
and thus the dynamic range in the convolved image of a complex source \citep{Labels}. Fizeau combination however becomes inefficient with highly diluted apertures, since the narrow interference
peak appearing in the image of a point source, at the center of the much broader envelope diffracted by the subapertures, contains a small proportion of the energy. A way of retrieving most
energy in the peak, for efficiently observing faint sources with many-aperture interferometers capable of rich direct imaging, has appeared in the form of the hypertelescope or "densified aperture"
scheme \citep{lab96,lar07}. Practical designs for large  hypertelescopes, with a spherical geometry inspired from the Arecibo radio telescope, are under testing for terrestrial versions
\citep{coroll11,rondi11,lund11,lab2012} and also studied for space versions \citep{LabESA,LaserBarcelo}  .
These  future large direct imaging interferometers  using many apertures will greatly benefit from adaptive optics systems for ultimate performance on faint sources,
providing usable imaging with large exposure time. But even in the absence of adaptive  phasing,  high resolution imaging can be done in speckle mode.
Speckle mode techniques like speckle interferometry \citep{lab70} and speckle masking \citep{wei77,loh83},  heretofore successfully used with large monolithic telescopes, can also produce useful results with such interferometers.

We describe the numerical simulations done to understand the scope and explore the performance of speckle imaging techniques with this type of diluted aperture interferometers.
We have built a numerical simulation code in MATLAB that simulates cophased and speckle mode imaging with diluted apertures in  different configurations.
The imaging performance of interferometers  is much affected by their beam combination scheme, as analyzed in detail by \cite{lar07}.
In our simulation we have adopted the pupil densification scheme \citep{lab96}, also adopted for the Ubaye Hypertelescope Project currently undergoing preliminary
testing in the Southern Alps \citep{lab2012,rondi11}. Pupil densification concentrates the diffracted energy in the central part of the Point Spread Function(PSF)  without degrading the resolution. A detailed
analysis of the PSF and imaging properties of these diluted aperture interferometers has been done by \cite{pat09} and \cite{lar07}. In the present article, our simulations
extend  these results to the case of randomly phased apertures exploited with speckle imaging, and also in the use of aperture rotation which enhances the quality of reconstructed images. We have also tried to find out the limiting magnitude for the technique by comparing Signal to Noise Ratio (SNR) of reconstructions at different 
photon levels.

\section[]{NUMERICAL SIMULATIONS}
According to the usual definition of hypertelescope imaging we consider a multi-aperture Fizeau interferometer equipped with a non-distorting pupil densifier i.e, one which
does not distort the pattern of sub-aperture centers.  This makes the interference function field-invariant, while the spread function depends on the star's position, through
its diffraction factor. This variable spread function,  hereafter called ``pseudo spread function'', results in a pseudo-convolution process which describes the image formation on extended sources. 
For $n_{T}$ identical sub-apertures with position vectors $(u_{k},v_{k}$), the cophased Pseudo Spread Function \citep{lab96}
is expressed as

\begin{figure}
\vspace{1mm}
\centering
\includegraphics[width=85mm , angle=0]{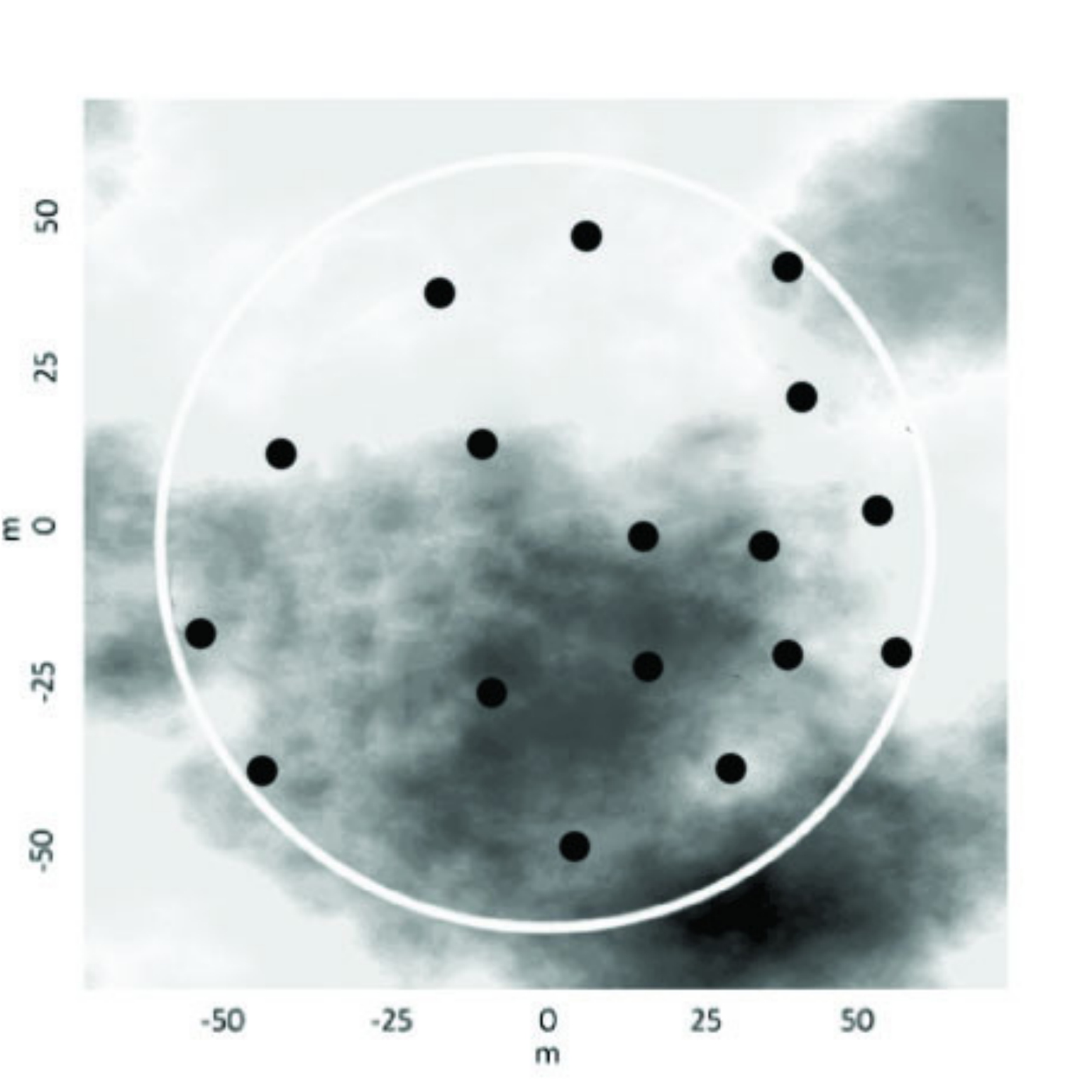}
\caption{ Random phase screen across the 20 mirror aperture used for the simulations, following a Kolmogorov distribution. The dots indicate positions of the subapertures in the fully diluted array.}
\label{fig1}
\end{figure}

\begin{figure}
\vspace{1mm}
\centering
\includegraphics[width=87mm , angle=0]{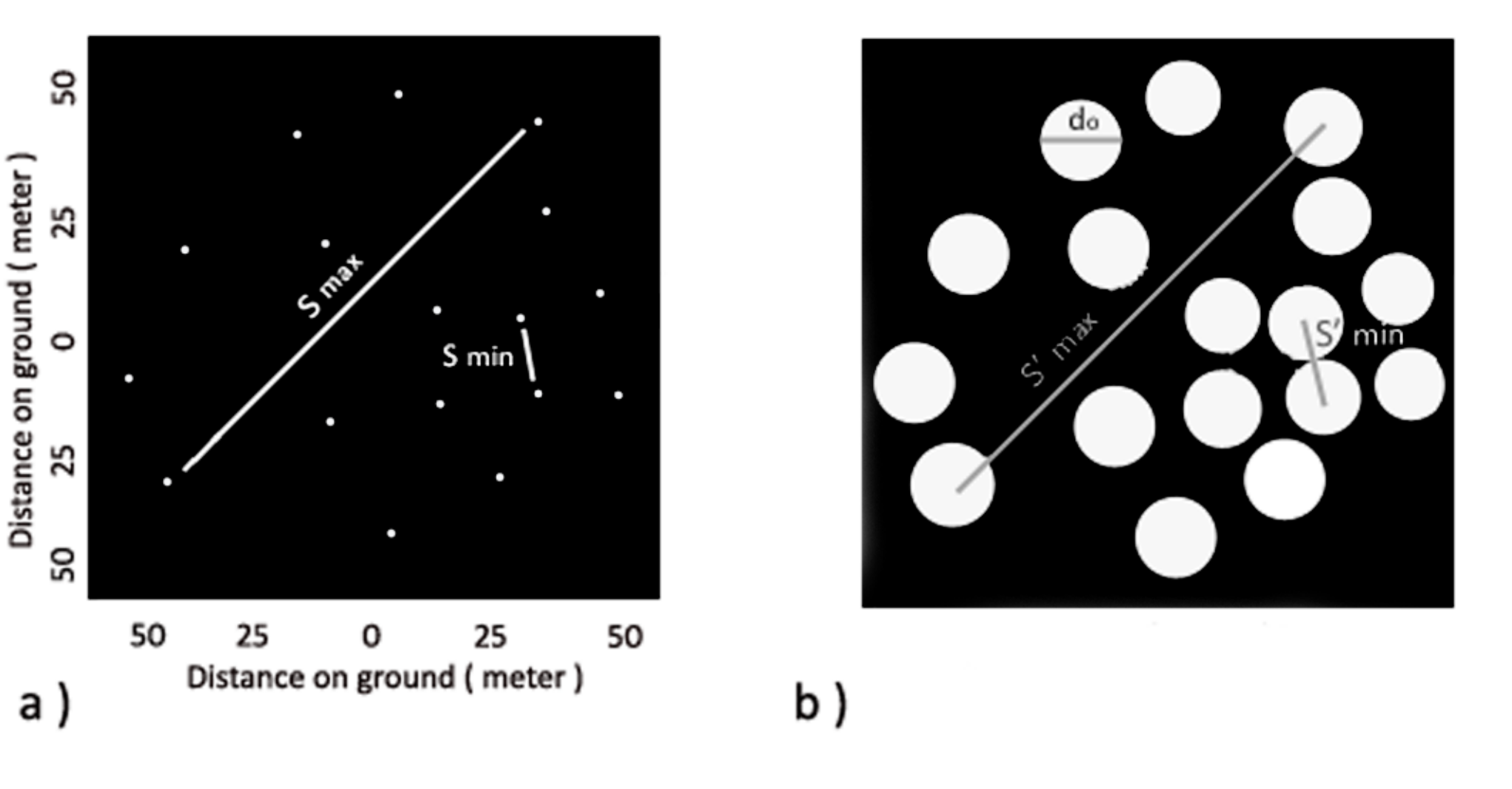}
\caption{Pupil densification of the aperture configuration used in the simulations. a) The input pupil of a diluted random 20 mirror array. b) The exit pupil after densification}
\label{fig2}
\end{figure}

\begin{figure}[h]
\centering
\includegraphics[width=85mm , angle=0]{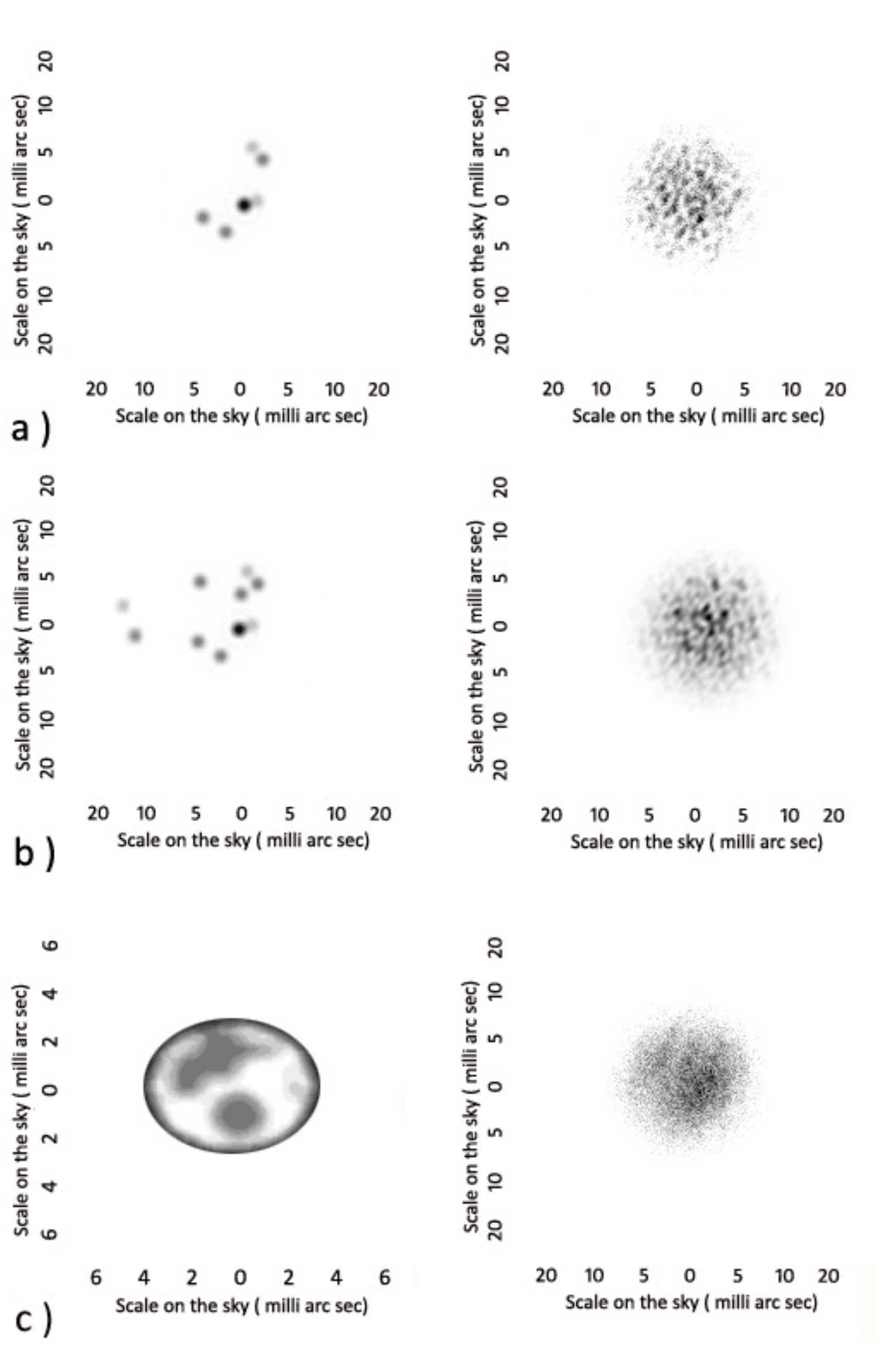}
\caption{The object distributions used in the simulations (first column) along with their corresponding speckle frames (second column) as imaged by the 20 mirror array. a) Six star group b) Ten star group c) Extended object.
Speckle frames in an average have 10000 photon events per each exposure.}
\label{fig3}
\end{figure}

\begin{equation}
\hspace{15 mm}
I_{PSF}(x,y)=A(x,y)\times I_{0}(x,y),
\end{equation}
where $A(x,y)$ is the diffraction function and the term on the right side, $I_{0}(x,y)$ is the Interference function given by

\begin{equation}
\hspace{15 mm}
I_{0}(x,y)=\left|\sum_{k=1}^{n_{T}}e^{-\frac{2\pi i}{\lambda}(xu_{k}+yv_{k})}\right|^{2}.
\end{equation}
The interference function depends only on the array configuration while the diffraction function $A(x,y)$ depends on the beam combination
scheme used \citep{lar07}. 
For our analysis we have considered $A(x,y)$ to be an Airy function corresponding to the densified subaperture diameter $d_0$. In our simulation we are considering highly diluted arrays, approximated to array of $\delta$-functions, in which pupil
densification is so strong that the shift of the Airy envelope is negligible. In such a case the image formation of the hypertelescope can be defined by a Pseudo Convolution Equation

\begin{equation}
\label{pseudoconv}
\hspace{15 mm}
I(x,y)= A(x,y)\times\left(I_{0}(x,y)\otimes O(x,y)\right),
\end{equation}
where $O(x,y)$ is the object intensity distribution. The pseudo convolution creates the Direct Imaging Field of the interferometer
which is the diffraction function envelope corresponding to the densified subaperture. This modeling of hypertelescope imaging is similar to the study by \cite{aime08}.

To simulate the seeing conditions we generate  Kolmogorov phase screens (fig. \ref{fig1})  with different  values of Fried parameter and also move these screens with
different wind velocities to simulate seeing changes. In our simulations we assume that the wavefront is coherent across an individual subaperture. This will be the case if you use small mirrors ( $ < 10 cm$) in a site with good seeing conditions, and hence only piston errors associated with each subaperture was considered. The random atmospheric piston error  $e^{i\phi_{k}}$ from the corresponding position of the sub-aperture in
the phase screen is introduced into the interference function. 
\begin{equation}
\label{randomint}
\hspace{15 mm}
I_{0}(x,y)=\left|\sum_{k=1}^{n_{T}}e^{-\frac{2\pi i}{\lambda}(xu_{k}+yv_{k})}.e^{i\phi_{k}}\right|^{2} 
\end{equation}
This new speckle interference function is used in the pseudo-convolution equation  (\ref{pseudoconv}) to simulate speckle observations. Some of the speckle images corresponding to sample objects used in the study are shown in figure \ref{fig3}.
If the phase across each subaperture is not coherent and there are phase pertubations across the individual mirrors, the diffraction envelope will break in to a speckle pattern. We have not done an estimate of this effect in the current study. 

\subsection{Aperture Configuration}

Though there are several aperture configurations under consideration for the big imaging interferometers, the one we studied was of randomly
distributed mirrors with non-redundant baselines. The number of diluted apertures, $n_{T}$ of 20 (fig \ref{fig2}) and 50 were considered for this paper.
 The details of aperture configuration used in the simulations are in table \ref{tab1}. Details of the pupil densification beam combination scheme used in the simulations are discussed in detail 
by \cite{lar07} and \cite{lab96}. Densification is constrained by the smallest baseline $S_{min}^{'}$ in the output pupil, since any further densification
will cause subapertures to overlap. Thus in maximum densification, length of the smallest baseline in output pupil will be equal to the densified subaperture 
diameter, i.e. $S_{min}^{'}=d_{o}$. In our simulation  for all the array configurations we have taken densified diameter of each diluted subaperture $d_{o}$ to 
be ${1}/{10}$ of the largest baseline in the output pupil. Output pupil which is partially filled creates a speckle halo surrounding the central
peak in cophased case. The dominance of this halo is dependent on the output pupil filling rate ($\tau_{o}$).

\begin{table}
\center\begin{tabular}{|l|c|c|}

& 20 mirror  & 50 mirror
\tabularnewline
\hline
 &  & \tabularnewline
Maximum Baseline & 100 m & 100 m\tabularnewline
Minimum Baseline & 21 m & 12 m\tabularnewline
Output Pupil Filling rate $\tau_{o}$ & 0.2 & 0.5\tabularnewline
Sub-aperture diameter & 10 cm & 10 cm\tabularnewline
\hline

\end{tabular}

\caption{Parameters associated with aperture configuration.}
\label{tab1}
\end{table}

\begin{table}
\center\begin{tabular}{|l|c|c|}
\hline 

 &  & \tabularnewline

Frames used in speckle masking & 500 \tabularnewline
Earth Rotation time simulated & 8 h \tabularnewline
Plate scale for each frame & 0.2 milli arc sec /pixel \tabularnewline
Wavelength & 550 nm \tabularnewline
Bandwidth &  88 \AA{} \tabularnewline
Fried Parameter  & 0.1 m \tabularnewline
Wind Velocity     & 10 m/s \tabularnewline
Latitude of the place  & $+60^0$ \tabularnewline
Declination of the source  & $+90^0$ \tabularnewline
\hline
\end{tabular}

\caption{Parameters used in the simulation of hypertelescope imaging in the presence of atmospheric turbulence}
\label{tab2}
\end{table}

\begin{figure*}
\centering
\includegraphics[width=13cm , angle=0]{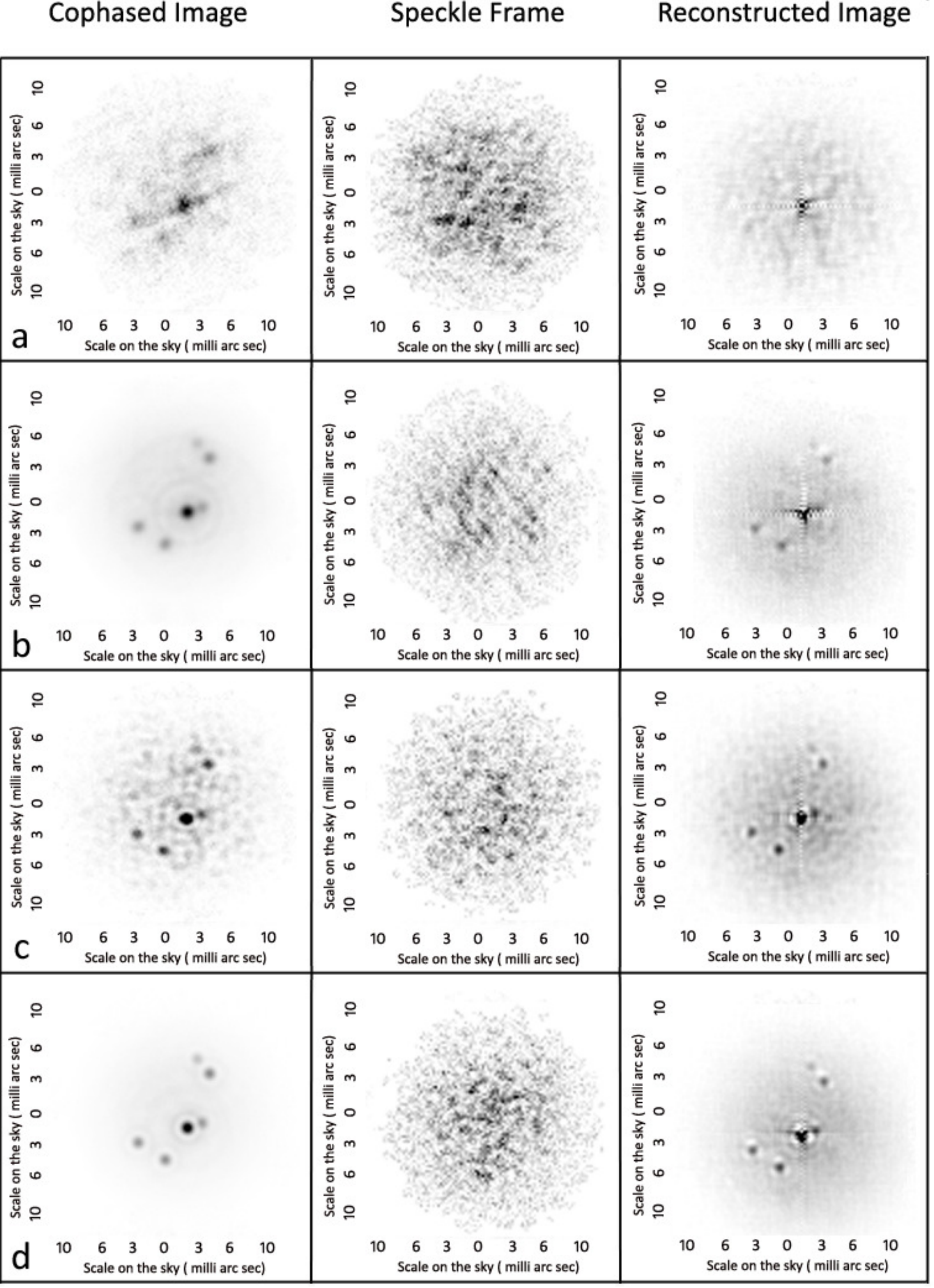}
\caption{Speckle imaging results for a six star cluster from turbulence degraded interferograms from 
a) 20 mirror aperture without aperture rotation, b) 20 mirror aperture with rotation through the night for 8 hrs,
c) 50 mirror aperture without rotation, d) 50 mirror aperture with aperture rotation through night for 8 hrs. Each simulated exposure on average had 10,000 photon events. }
\label{fig4}
\end{figure*}

\begin{figure}
\vspace{1mm}
\centering
\includegraphics[width=85mm , angle=0]{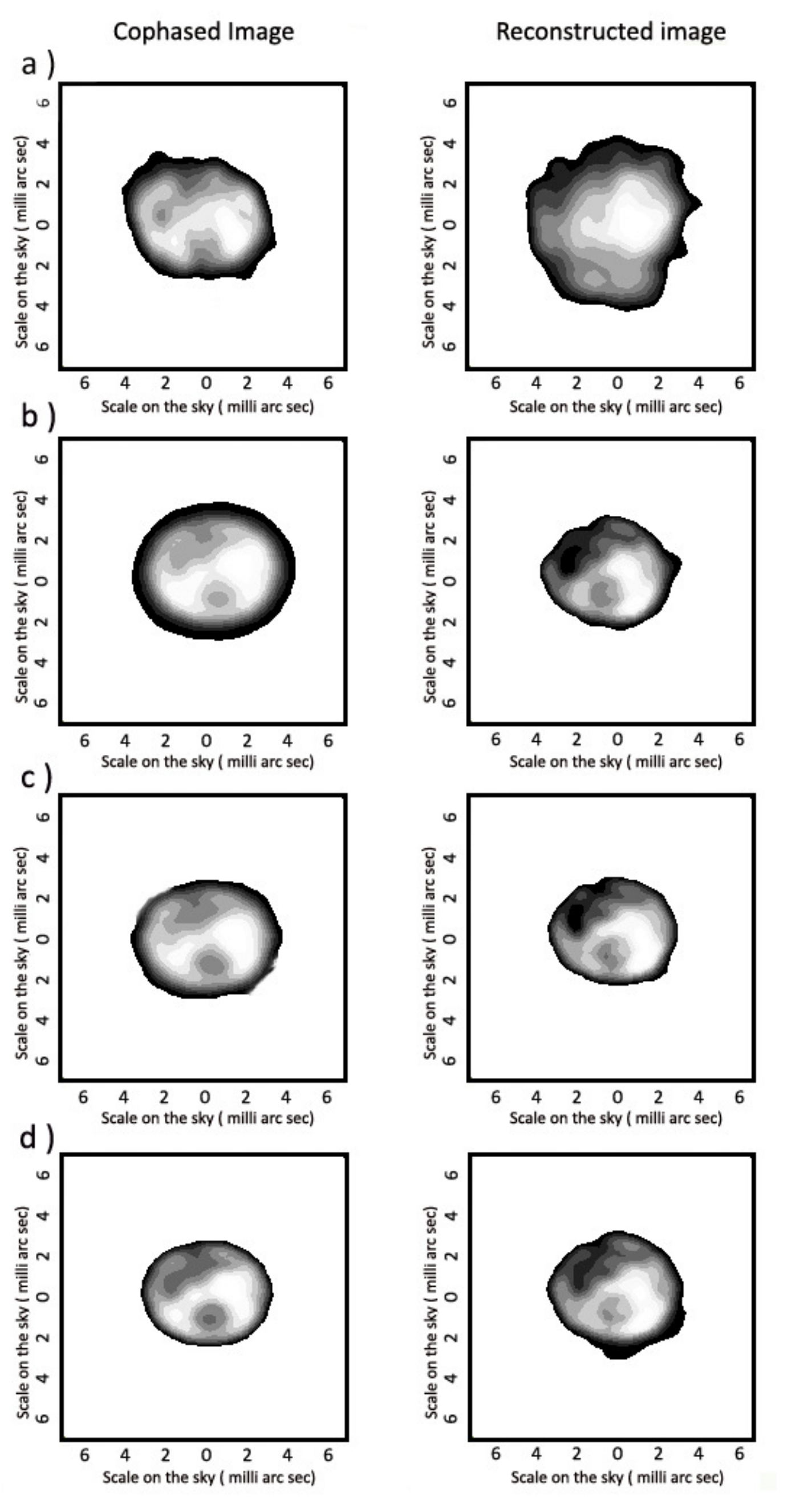}
\caption{Speckle imaging results for an extended object from turbulence degraded interferograms from a) 20 mirror aperture without aperture rotation,
b) 20 mirror aperture with rotation through the night for 8 hrs,
c) 50 mirror aperture without rotation, d) 50 mirror aperture with aperture rotation through night for 8 hrs. Each simulated exposure on average had 10,000 photon events.}
\label{fig5}
\end{figure}

\begin{figure*}
\centering
\includegraphics[width=13cm , angle=0]{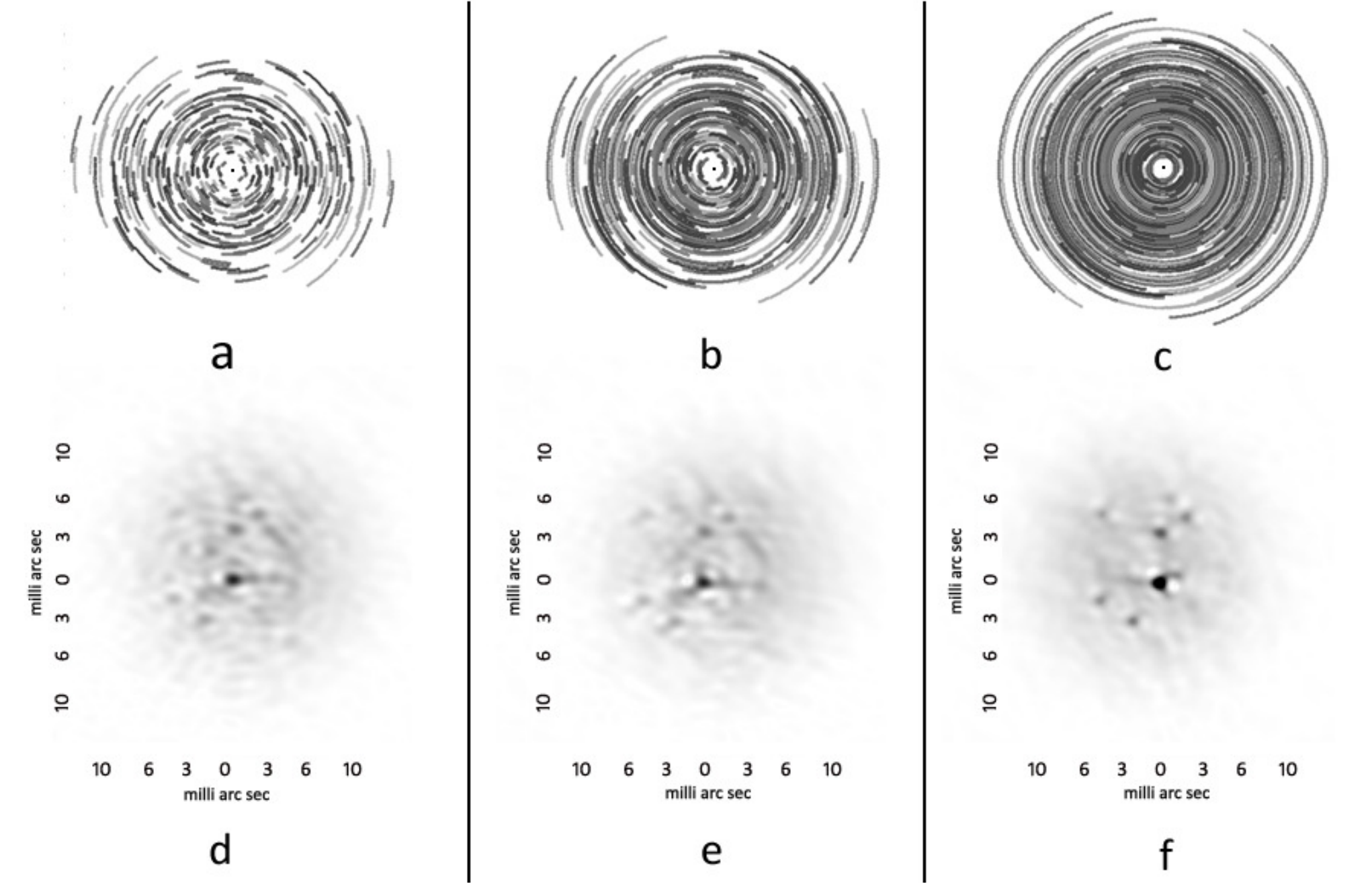}
\caption{Improvement of reconstructed image of star cluster with ten stars (Figure-\ref{fig2}.b), utilizing aperture rotation through night. a) u-v coverage of 20 mirror array over 1 hr of observation b) u-v coverage of 20 mirror
array over 3 hr of observation c) u-v coverage of 20 mirror array over 8 hr of observation.  d,e and f are the corresponding reconstructed images obtained by speckle imaging in each case.
The latitude of the site was taken as $60^0$ North and the declination of the star cluster is considered to be $90^0$. The images where reconstructed from 500 speckle frames each with an average of 10,000 photon counts.}
\label{fig6}
\end{figure*}

\subsection{Triple Correlation and Bispectrum}
The triple correlation technique developed by \cite{loh83} runs as 
follows. The object speckle pattern, $I(\bf{x})$, is 
multiplied with an appropriately shifted version of it, i.e., $I(\bf{x} + \bf{x_1})$. 
The result is then correlated with $I(\bf{x})$. 
\begin{equation}
I^{(3)}(\mathbf{x_1},\mathbf{x_2}) = \langle \int^{+\infty}_{-\infty} I (\mathbf{x}) I(\mathbf{x}+\mathbf{x_1})I(\mathbf{x}+\mathbf{x_2})dx \rangle 
\end{equation}
where, $\mathbf{x_j} = x_{jx} + x_{jy}$ is the 2-dimensional spatial 
co-ordinate vector. $\langle \rangle$ stands for ensemble average.

The Fourier transform of the triple correlation is called bispectrum and its
ensemble average is given by

\begin{equation}
\langle\widehat{  I}^{(3)}(\mathbf{u_1},\mathbf{u_2})\rangle = \langle\widehat I(\mathbf{u_1}) 
\widehat I^\ast 
(\mathbf{u_1} + \mathbf{u_2}) \widehat I(\bf{u_2})\rangle 
\end{equation}
where

$\widehat I(\mathbf{u}) =\int I(\mathbf{x})e^{-i2\pi \mathbf{u}.\mathbf{x}} d\mathbf{x}$,

$\widehat I^\ast(\mathbf{u_1} + \mathbf{u_2}) = \int I(\mathbf{x}) e^{i2\pi(\mathbf{u_1} + \mathbf{u_2}).\mathbf{x}}d\mathbf{x}$

and $\mathbf{u_j} = u_{jx} + u_{jy}$ is the 2-dimensional spatial frequency vector.

In the second order moment phase of the object's Fourier transform is lost,
but in the third order moment or in the bispectrum it is preserved. The argument
of equation (6) can be expressed as
\begin{equation}
arg [\widehat I^{(3)}(\mathbf{u_1}, \mathbf{u_2})] = 
\phi_b(\mathbf{u_1}, \mathbf{u_2}) = \phi(\mathbf{u_1}) - \phi(\mathbf{u_1} + \mathbf{u_2}) + 
\phi(\mathbf{u_2}) .
\end{equation}
Equation (7) gives the phase of the bispectrum. The phase values of the averaged image bispectrum are equal to that of the object bispectrum. This allows for the opportunity to extract real phase information from the 
object bispectrum. The modulus $\mid\widehat O(\bf{u})\mid$ and phase $\phi(\bf{u})$ of the 
object Fourier transform $\widehat O(\bf{u})$ can be derived from the object 
bispectrum $\widehat I_{O}^{(3)}(\bf{u_1},\bf{u_2})$.  
The object phase-spectrum is thus encoded in the term
$e^{i\phi_b(\bf{u_1},\bf{u_2})} = e^{i[\phi(\bf{u_1}) 
- \phi(\bf{u_1} + \bf{u_2}) + \phi(\bf{u_2})]}$ of the 
equation (6).

Equation (7) is a recursive equation for evaluating the phase of the object 
Fourier transform at coordinate $ \bf{u} = u_1 + u_2$. 
The phase of the
bispectrum is recursive in nature and the object phase-spectrum at $(\bf{u_1}
+ \bf{u_2})$ can be expressed as

\begin{equation} 
\phi(\bf{u_1} + \bf{u_2}) = \phi(\bf{u_1}) + \phi(\bf{u_2}) - \phi_b(\bf{u_1},\bf{u_2}) .
\end{equation}

This recursive technique is used in the algorithm to retrieve the object Fourier phase from the Bispectrum phase. 

\subsection{Speckle Imaging Algorithms}

For the speckle images generated from numerical simulation of hypertelescopes, techniques like speckle interferometry and bispectrum technique were applied to reconstruct
the object distribution. The speckle interferometry code was written in MATLAB and produces the Fourier amplitude information and autocorrelation of the object distribution.
In speckle interferometry, the average power spectrum $\sum|I(u,v)|^{2}$ is found out from
the speckle frames and object Fourier amplitude information is extracted from it. The computed Fourier amplitude information is used in speckle masking. 
Bispectrum based reconstruction code was also written in MATLAB by one of us (AS).
Though a computationally efficient tomographic speckle masking (TSM) code \cite{Suryall}, has also been developed by AS, we have for the current study used the direct bispectrum code which gives a better quality of reconstruction because it uses the four dimensional bispectrum. The direct bispectrum code can process 200 x 200 pixels images of 300 frames in 15 minutes in a i7 Intel computer with 8 GB of RAM. The unit amplitude phasor method,
used in algorithms by \cite{sri00}, is used in the code for phase reconstruction. The details of the reconstruction code is explained in \cite{Suryall}. 
The code uses direct computation of  the 4-dimensional bispectrum $I^{(3)}(u,v,u^{'},v^{'})$ which
poses severe constraints on the computer memory. The 4-D bispectrum is computed and averaged out for all the speckle frames. The Fourier phase is retrieved from the bispectrum using the techniques explained in section 2.2, and is combined with Fourier amplitude from speckle interferometry to reconstruct the object.

\subsection{Earth Rotation Aperture Synthesis}

The sparse filling of the entrance aperture, although enhanced in the densified exit pupil, affects the performance of the speckle imaging reconstruction.
But part of it is retrieved if the entrance aperture, as seen from the observed star, can be modified or rotated during an observation. This happens naturally for interferometers of fixed ground elements
due to earth's rotation. Earth Rotation Aperture Synthesis is a common technique frequently exploited in radio interferometry to increase the coverage of the frequency plane by an interferometric array.
With the help of simulations we have tried to study the possible techniques of using aperture rotation through night with diluted aperture hypertelescope systems.
\cite{rein93,rein97} had earlier studied the use of aperture rotation with speckle masking for studies of LBT and VLTI interferometers in multi speckle mode.
Our simulations address the use of such techniques with long baseline hypertelescope arrays. 

Using speckle frames from different time of the night averaged power
spectrum, $\sum_{i}|I(u,v,t_{i})|^{2}$, provides a better u-v coverage and thus a better estimate of the Fourier modulus to be used in reconstruction.
When speckle frames from different time of the night were used, the time dependent bispectrum, $I^{(3)}(u,v,u^{'},v^{'},t_{i})$,
of each frame was computed and averaged. From the averaged bispectrum, $\sum_{i}I^{(3)}(u,v,u^{'},v^{'},t_{i})$, the image of the object was reconstructed. The averaged bispectrum contains the u-v coverage
as sampled through the night by earth rotation and hence provides a better reconstruction of object phase.

\section{Results}

%

The numerical simulations provide a clear picture of how speckle masking can be used together with hypertelescope imaging utilizing aperture rotation to yield high resolution images of stellar objects. The bispectrum code was utilised to reconstruct images from the speckle images simulated from the diluted hypertelescope. We have quantified the reconstruction quality using correlation coefficient $c$ which measures the correlation between the reconstructed image and the cophased image in the absence of atmospheric turbulence. The correlation coefficient, $c$ is computed according to the following equation

\begin{equation}
c=\frac{\sum_m\sum_n (A_{mn}-\bar{A})(B_{mn}-\bar{B})}{\sqrt{(\sum_m\sum_n (A_{mn}-\bar{A})^2)(\sum_m\sum_n (B_{mn}-\bar{B})^2)}},
\end{equation}
where $A$ and $B$ are the true image and the reconstructed image respectievely, $\bar{A}$ and $\bar{B}$ are the mean pixel count and ($m$,$n$) are the pixel positions.

Such quantification of reconstruction gives us the ability to compare the reconstructions with different parameters.

\subsection{Reconstruction results with sextuple star and extended objects}
We have simulated sequences of short exposures, shorter than the lifetime of "seeing", and each exploiting the full set of sub-apertures. The parameters associated with this simulation are shown in table \ref{tab2}.  Two different sequencing regimes were simulated:   
a) dense sequences, with 1000 short exposures  made in a matter of minutes while the slow Earth rotation causes a negligible rotation of the meta-aperture with respect to celestial North;   
b) night-long sequences, where Earth rotation becomes significant and exploitable for aperture-supersynthesis.  
The speckle images from the simulations were processed by the bispectrum technique to obtain reconstructions. The results of these simulations are shown in Figure~\ref{fig4} and \ref{fig5}. 
When available, long sequences thus exploited  improve the result quality , especially if there are a small number of sub-apertures . The correlation coefficient $c$ between reconstructed image and the corrosponding cophased image for both the objects with different arrays are shown in table \ref{tab3}.

\subsection{Improvement of Reconstruction with Aperture Rotation}

The improvement of signal reconstruction in speckle masking using aperture rotation is clearly demonstrated in figure~\ref{fig6}. The object used is a star cluster with 10 stars (figure \ref{fig3}.b). The aperture used is a 20 mirror array with rotation synthesis corrsponding to 1 hour, 3 hour and 8 hour. The respective correlation coefficient for each of the reconstructions is shown in Table 4.  We have correlated the reconstructed image here with the true object image. The improvement in $c$ shows the  improvement of the reconstruction with better spatial frequency coverage.   It is also curious to note the field of view limitation due to pupil densification as also seen from the figure~\ref{fig6}. Only 8 stars out of the 10, which are inside the
Direct Imaging Field are seen from the reconstructed image.

\begin{table}
\center\begin{tabular}{|l|c|c|c|}

&Aperture Rotation& 20 mirror  & 50 mirror
\tabularnewline
\hline
 &  & \tabularnewline
\multirow{2}{*}{Sextuple Star}& No& .60 & .71 \\
 & Yes & .77 & .78 \\ \hline
 \multirow{2}{*}{Extended Object}& No  & 0.5 & 0.90 \\
 & Yes& 0.88 & 0.92\\ 

\tabularnewline
\hline

\end{tabular}

\caption{The values of correlation coefficient $c$ corresponding to the reconstruction of sextuple star and extended object from the speckle images with 20 mirror and 50 mirror array. }
\label{tab3}
\end{table}

\begin{table}
\center\begin{tabular}{|l|c|c|c|}

& 3 hour  & 6 hour & 8 hour
\tabularnewline
\hline
 &  & & \tabularnewline
Correlation coefficient $c$ & 0.49 & 0.52 & 0.62\tabularnewline

\hline

\end{tabular}

\caption{The correlation coefficient values corresponding to the reconstruction of the star cluster with increasing hours of observation to utilize aperture rotation. The array used is of 20 sub-apertures.}
\label{tab4}
\end{table}

\subsection{Limiting Magnitude}
\begin{figure}
\vspace{1mm}
\centering
\includegraphics[width=85mm , angle=0]{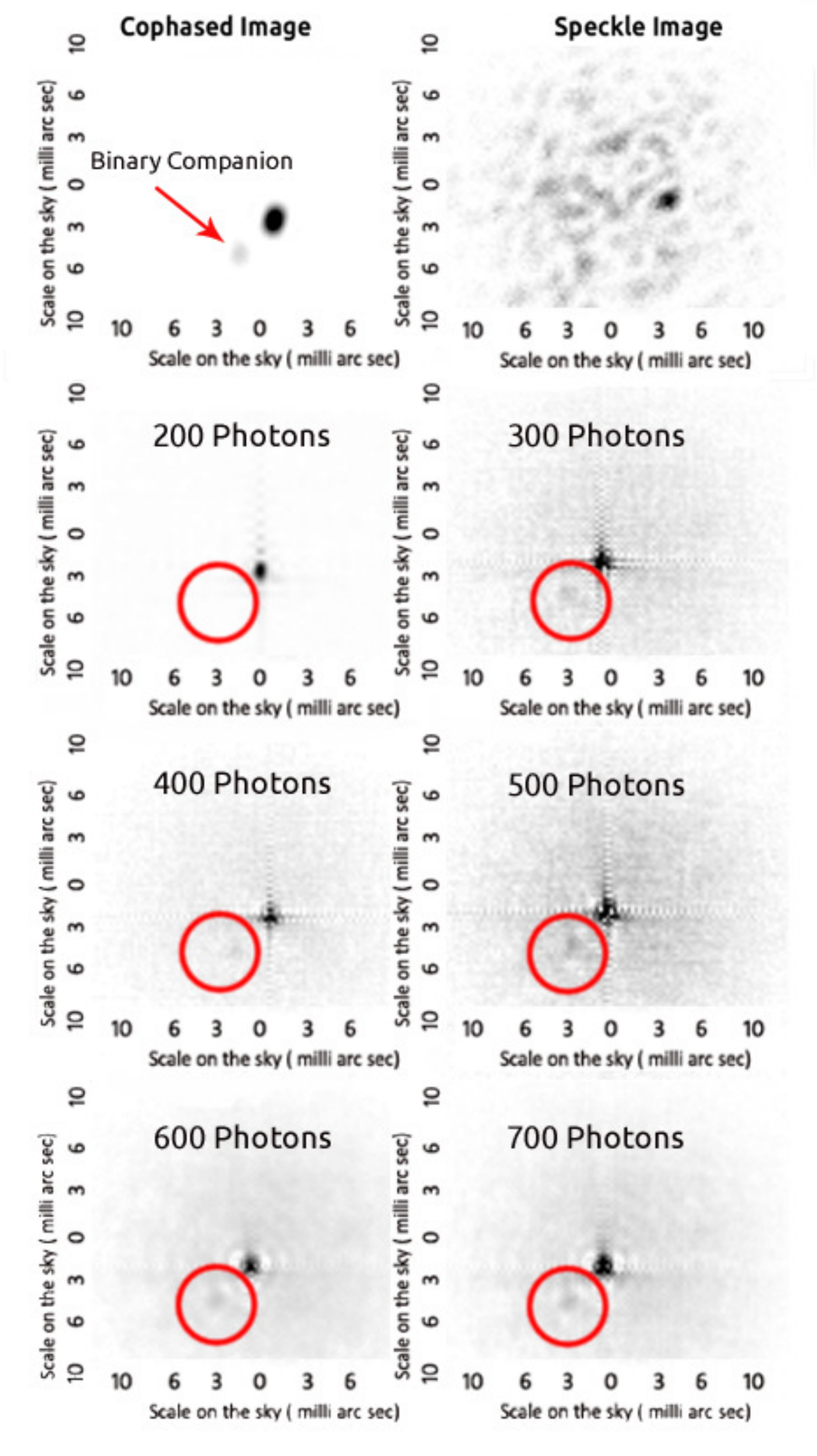}
\caption{Reconstruction of a Binary Star with Brightness ratio 1000:1 at different magnitudes. The aperture used is an array of 50 mirrors with 7 hours of earth roatation aperture synthesis, which increases the spatial frequency coverage. (Top row) The ideal cophased image and speckle image of the binary star. (Bottom three rows) The reconstructed images using the bispectrum technique with the corrosponding photon counts. The region inside the red circle is used to compute the SNR.  }
\label{fig7}
\end{figure}

\begin{figure}
\vspace{1mm}
\centering
\includegraphics[width=90mm , angle=0]{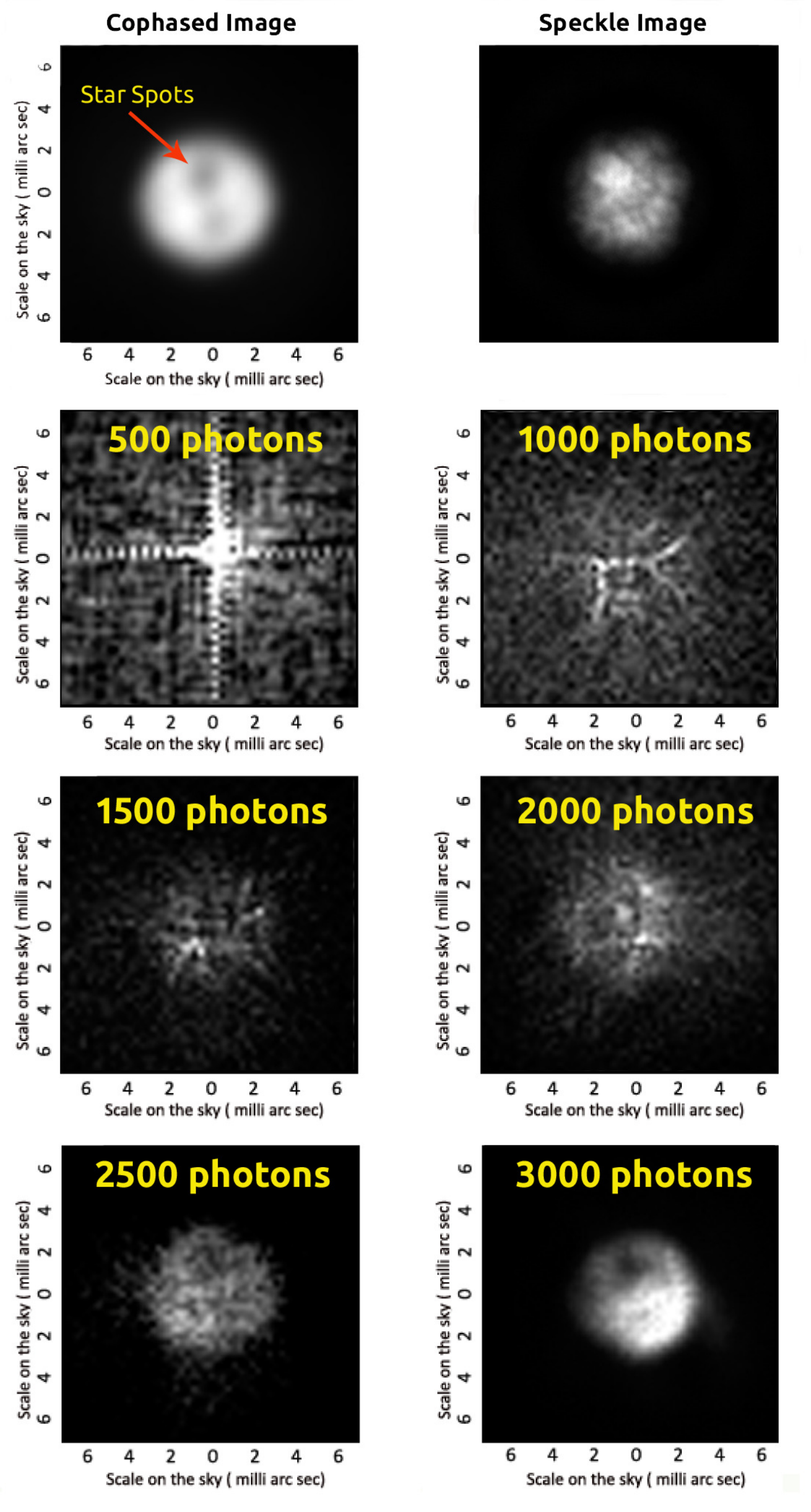}
\caption{Reconstructions of a resolved star with spots at different magnitudes.The aperture used is an array of 50 mirrors with 7 hours of earth roatation aperture synthesis, which increases the spatial frequency coverage. (Top row) The ideal cophased image and speckle image of the resolved star. (Bottom three rows) The reconstructed images using the bispectrum technique with the corrosponding photon counts.}
\label{fig8}
\end{figure}

We wanted to compute the limiting magnitude of objects retrievable through such a technique. For this we have simulated imaging of two objects, a binary star and a resolved star with spots. Both the objects were imaged assuming different magnitudes progressively and the Signal to Noise Ratio was computed for each recontruction. Since the photon levels of speckle images used for reconstructions to compute limiting  magnitude were very low, we have used another measure to compute the Signal to Noise Ratio instead of the correlation coefficient.  The SNR for the binary star was computed as the contrast of the fainter star in the background halo of noise.

\begin{equation}
SNR= \frac{\mu_{star}}{\sigma_{background}}  
\end{equation}

where $\mu_{star}$ is the mean flux level from the pixel positions corresponding to the star and $\sigma_{background}$ the standard deviation of the average background noise.

\begin{figure}
\vspace{1mm}
\centering
\includegraphics[width=80mm , angle=0]{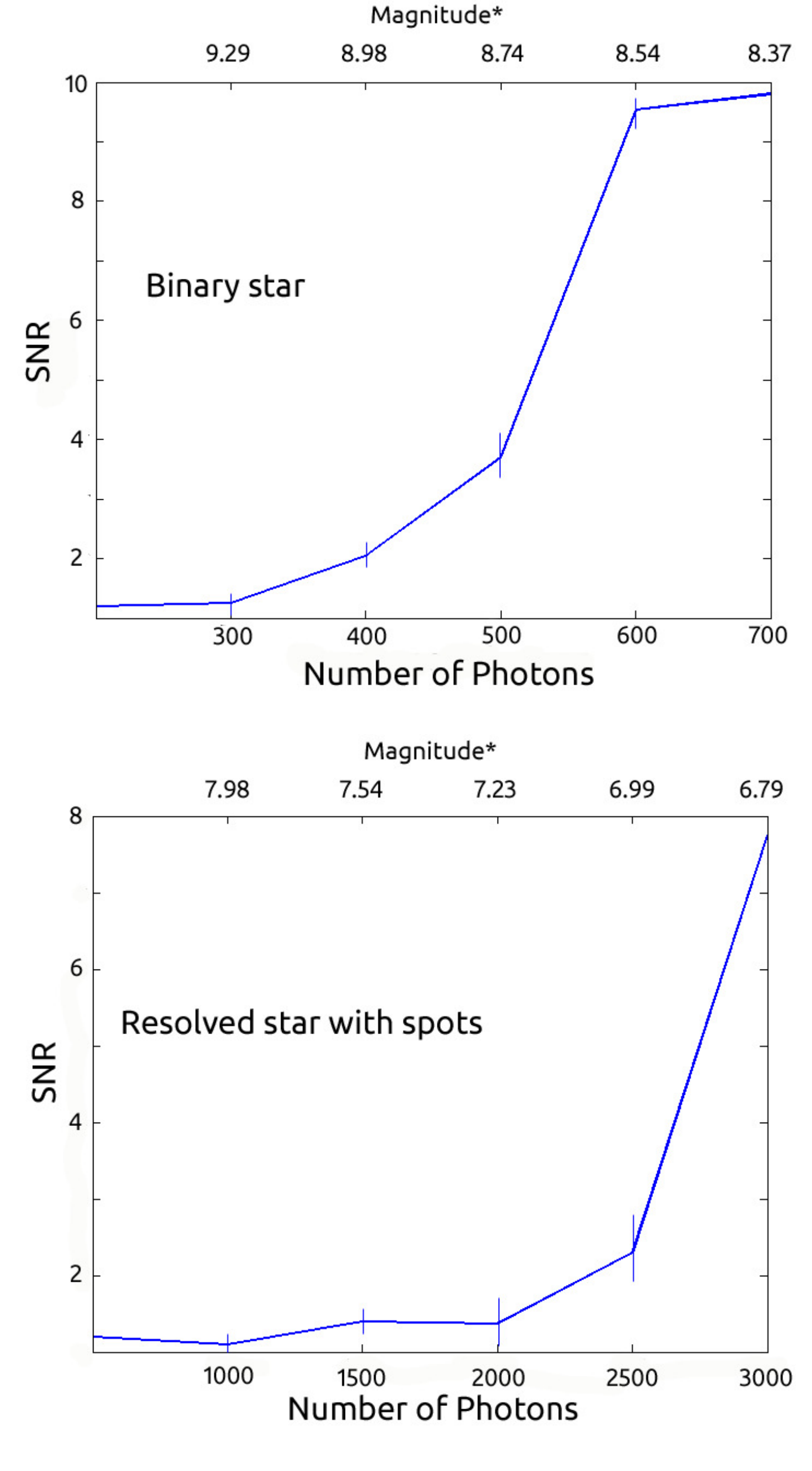}
\caption{ The SNR of the reconstructed images plotted against photon count in the speckle images for both Binary star and the resolved star with spots.
*:   The magnitude is calculated corresponding to the parameters used in the simulation as listed in table \ref{tab1} and \ref{tab2}.}
\label{fig9}
\end{figure}

The average noise was computed in the encircled area in the figure (\ref{fig7}). For the star spots the contrast level of one of the spots was used as the measure of SNR.  To calculate the SNR, 20 reconstructions were computed at each magnitude photon level.
The resulting SNR plotted against the photon levels are shown in figure \ref{fig9} for both binary star and the resolved spotted star. 

\section{Conclusions}
The simulations have successfully shown that even in the absence of adaptive phasing, hypertelescope systems employing pupil densification can be used for direct imaging using speckle imaging
techniques. It is shown that with utilizing the aperture rotation of the diluted array through night, reconstruction quality could be increased substantially.  Other ways
of changing the aperture pattern during observation may also be considered, and should similarly be expected to improve the imaging performance.
Though as with cophased imaging, speckle imaging also is constrained by the field of view limitations in a hypertelescope. The use of speckle imaging for future Earth-based hypertelescopes \citep{lab2012b} can be of interest  for faint sources  which cannot be phased in the absence of
adaptive optics, or because of the absence of a sufficiently bright guide star. 
With the parameters of hypertelescope used in the simulation we have obtained a good SNR for the speckle technique at a magnitude of 8-9 for a simple binary star and 6-7 for a resolved star with spots.
The improvement in limiting magnitude of the technique with different amounts of pupil densification need to be studied further.
Also the technique could give better results with deconvolution algorithms and pupil re-dilution as studied by \cite{Aime2}.
While modified forms of adaptive optics have been proposed specifically for hypertelescopes  \citep{Borko,Martinache},
it  remains unclear at this stage whether suitable forms of a Laser Guide Star can also be operated for such instruments \citep{Paul}.
If not, Earth-based hypertelescopes will have to use speckle-imaging on faint sources, which should provide useful results, according to our simulation results,
although with lesser performance  than in phased conditions. For space hypertelescopes, phasing is expected to be much easier \citep{LabESA,LaserBarcelo}. Although  the huge sizes considered,
100 km for an Exo-Earth Imager and 100,000 km for a Neutron Star Imager, can make them sensitive to new forms of seeing such as ``gravitational seeing" .

\label{lastpage}

\end{document}